\documentclass[twocolumn,prb]{revtex4-1}% Physical Review Letters

\usepackage[]{graphicx}

\begin{document}

\title{Interplay of electron-phonon nonadiabaticity and Raman scattering \\
  in the wavepacket dynamics of electron-phonon-photon systems}
\author{Kunio Ishida}
\email{ishd_kn@cc.utsunomiya-u.ac.jp}
\affiliation{Department of Electrical and Electronic Engineering, School of Engineering\\
  and Center for Optical Research and Education, Utsunomiya University\\
7-1-2 Yoto, Utsunomiya, Tochigi 321-8585, Japan}

\begin{abstract}
Ultrafast wavepacket dynamics of electron-phonon-photon systems is studied by numerical calculations. 
When nonadiabaticity of electron-phonon systems is taken into account, Raman scattering process plays an important role in the dynamics of the system.
We found that the interplay of the electron-phonon nonadiabaticity and the Raman scattering determines the wavepacket motion particularly in the vicinity of the conical intersection of adiabatic potential energy surfaces, which shows that we should consider this effect in order to reveal the photoexcitation/deexcitation process of materials in femtosecond time scale.
\end{abstract}

\maketitle

\section{Introduction}

Recent progress of ultrashort pulse laser technology has made it possible to observe time evolution of quantum-mechanical states in coherent regime.
Above all, time-resolved techniques have been developed to clarify the excited-state dynamics of condensed matter, and its transient properties have been studied\cite{measure1,measure2,measure3,measure4}.
When, for example, ultrashort laser pulses are used as a probe, transient electronic properties are detected through the slight change of optical properties, which helps us understand the physics in ultrafast timescale.
On the other hand, since the structural properties are also of interest in studying relaxation dynamics, time-resolved x-ray or electron diffraction measurement has been intensively developed, and the dynamics of photoinduced structural change or chemical reaction is currently being discussed by many authors\cite{pp1,pp2}.
Furthermore, as various experimental methods have been developed to understand transient phenomena\cite{pp3,pp4}, it is quite important to combine as many experimental data as possible in order to obtain an overall picture for the transient dynamics of certain materials.
We, however, should point out that we need a unified physical model to understand all the experimental data on a single phenomenon.
In particular, when we are interested in time evolution of excited states, we consider that a first step to construct an appropriate model is an accurate description of photoexcitation/deexcitation processes.

On the other hand, coherent control of quantum-mechanical states have been studied since the proposal of the adaptive control method\cite{rabitz}, which are considered to be promising for innovative technology particularly in the field of quantum information.
For example, recent experiments on diamond showed the possibility of phonon-mediated coherence between qubits via Raman scattering processes\cite{sussman}.
In this case, entanglement between electrons, phonons, and photons plays an important role, which means that theory of quantum entanglement between irradiated light and materials is necessary to reveal and/or design the control methods for them.
In many of theoretical studies, however, electromagnetic field has been regarded as classical external field, and only the physical variables on the material side are quantized.
We note that, when we are interested in the interplay among electrons, photons, and phonons, we should study the quantum dynamics of the full Hamiltonian regarding all the above entities as quantum variables.

In this paper, we study the quantum dynamics of electronic systems under photoirradiation, focusing on the wavepacket states created by photons.
We consider that we will have a reliable information on the response to various probes and the entanglement properties between multiple degrees of freedom by determining the quantum nature of those created wavepackets.
Since, however, it is not straightforward to obtain a first-principles theory for transient dynamics, we focus on the interplay of electron-phonon interaction and electron-photon (electric dipole) interaction as a first step to construct a general theory of transient photoexcitation/deexciation phenomena.
For this purpose we chose a model of coupled electron-phonon-photon systems and solved the time-dependent Schr\"odinger equation for fully quantized systems in order to discuss the wavepacket motion in the presence of electromagnetic field.

We also mention that nonadiabatic coupling between potential energy surfaces (PESs) is a key to understand the relaxation dynamics of photoexcited states, e.g., photoinduced nucleation\cite{me1}.
Furthermore, when electrons, phonons, and photons are considered at the same time, the Raman processes are expected to give significant contribution to the electronic transitions, which means that the photoexcitation/deexcitation process of electron-phonon systems should be carefully dealt with in order to understand the ultrafast dynamics with light-matter interaction.
As well as the intermodal coherence mediated by the Raman processes\cite{sen}, we stress that it is worth while mentioning that previous studies\cite{mukamel,ci2,citheory,xie} have shown that the conical intersection(CI) in the ``classical'' adiabatic PESs also is a key to understand the wavepacket dynamics.
These results show us that the coexistence of nonadiabatic coupling between PESs and Raman scattering processes gives another viewpoint on the coherent dynamics of electron-phonon-photon systems.
The aim of the present paper is to study the interplay of electron-phonon nonadiabaticity and Raman scattering processes by numerical calculations with a toy model.

The paper is organized as follows: the model and the calculation method are introduced in Sec.\ \ref{mm}, and the calculated results are presented and discussed in Sec.\ \ref{results}.
Section \ref{concl} is devoted to summary and conclusions.

\section{Model and method}
\label{mm}
In this paper we study the quantum dynamics of electrons coupled with both phonons and photons taking into account the Raman scattering processes.
For this purpose, we employed a model of a two-level electronic system coupled with a single-mode phonons and multimode photons described by
\begin{eqnarray}
{\cal H} & = &\omega a^\dagger a + \sum_{i=1}^n \Omega_i c_i^\dagger c_i +   \sigma_x \{ \sum_{i=1}^n \mu_i (c_i^\dagger + c_i ) + \lambda \} \nonumber \\
& + & \frac{1}{2}(\sigma_z +1)\{ \nu (a^\dagger + a) + \varepsilon \},
\label{ham}
\end{eqnarray}
where $a$ and $c_i$ denote the annihilation operators of phonons and the photons of the $i$-th mode, respectively.
$\sigma_i$ corresponds to the Pauli matrices which operate on the electronic states denoted by $|g \rangle$(ground state) and $|e \rangle$(excited state).
We should refer to the Jaynes-Cummings model\cite{jc} which formally includes the same type of interaction between electrons and bosons as Eq.\ (\ref{ham}), although we consider two kinds of bosons, i.e., phonons and photons.

Corresponding PESs of the Hamiltonian (\ref{ham}) are obtained by regarding the amplitude operators $\hat{u}=(a^\dagger + a)/\sqrt{2\omega}$, $\hat{v}_i = (c_i^\dagger + c_i)/\sqrt{2\Omega_i}$ as classical variables.
Although we do not consider any classical motion on those PESs, it still helps us understand the overall behavior of the wavepackets by discussing the adiabatic PESs which are given by
\begin{eqnarray}
U_\pm(u,v_1, v_2,..., v_n) & = & \frac{\omega^2}{2} u^2 + \sum_{i=1}^n \frac{\Omega_i^2}{2} v_i^2 + \frac{1}{2} (\nu' u + \varepsilon) \nonumber \\
 & \pm & \sqrt{(\nu u' + \varepsilon)^2+(\sum_{i=1}^n \mu'_i v_i + \lambda)^2},
\label{pes}
\end{eqnarray}
where $\nu' = \sqrt{2\omega} \nu$ and $\mu'_i = \sqrt{2\Omega_i} \mu_i$.
Equation (\ref{pes}) shows that the adiabatic PESs have a CI given by $u = -\varepsilon/\nu'$ and $\sum_i^n \mu'_i v_i = -\lambda$, and thus the geometrical phase of the wavefunction plays an important role on the dynamical properties\cite{citheory}.

The time-dependent Schr\"odinger equation for Hamiltonian (\ref{ham}) is numerically solved for $n=1$ and 3 and to obtain the wavefunction $|\Phi(t)\rangle$.
The initial condition is given by $|\Phi(0) \rangle = |\alpha_1, \alpha_2,\cdots, \alpha_n\rangle \otimes |0 g\rangle$, where $|\alpha_i \rangle$ denotes a coherent state parametrized by $\alpha_i$ for the $i$-th photon mode and $|0 g \rangle$ is the ground state of the electron-phonon system.
The values of the parameters are $\omega=1$, $\mu_i=0.5 (i=1,2,3)$, $\nu=3.5$, and $\varepsilon=13.5$, which shows that the electron-phonon coupling (Huang-Rhys factor) has intermediate strength between solid\cite{inorg} and typical organic molecules\cite{org}. 
As for the photons, we consider cases with a single mode ($n=1$) and three modes ($n=3$).
Mode 1 (pump mode) which is in resonance with the Franck-Condon transition is treated in both cases ($\Omega_1=13.5$).
Mode 2 (Stokes mode; $\Omega_2=12.5$) and mode 3 (anti-Stokes mode; $\Omega_3=14.5$) are taken into account in the latter case.

While the internal vibration mode of the material system is always quantized in the present study, we also calculated the dynamics of the material system treating the electromagnetic field as a classical external field for reference, and call this method a semiclassical approximation in the rest of the paper.

\section{Calculated Results}
\label{results}
In order to study the dynamical behavior of electrons, phonons, and photons, we first calculated the following properties: population of the electronic ground state $N(t) = \langle \Phi (t) | (1+\sigma_z)/2 | \Phi(t) \rangle$ and photon number in each mode $n_i(t)=\langle \Phi(t)|c^\dagger_i c_i|\Phi(t)\rangle$.

\begin{figure}
  \scalebox{0.5}{\includegraphics*{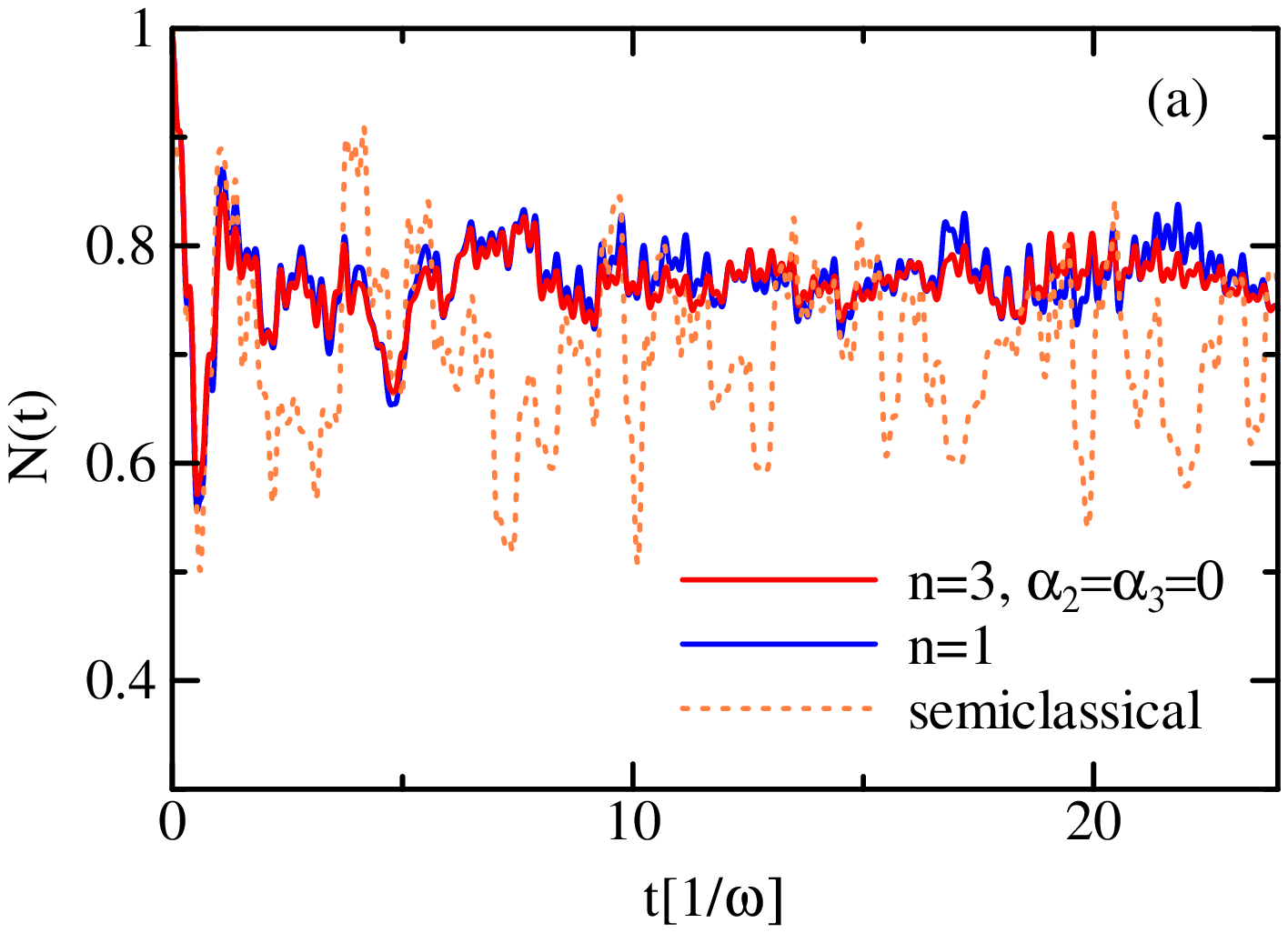}}
  \scalebox{0.5}{\includegraphics*{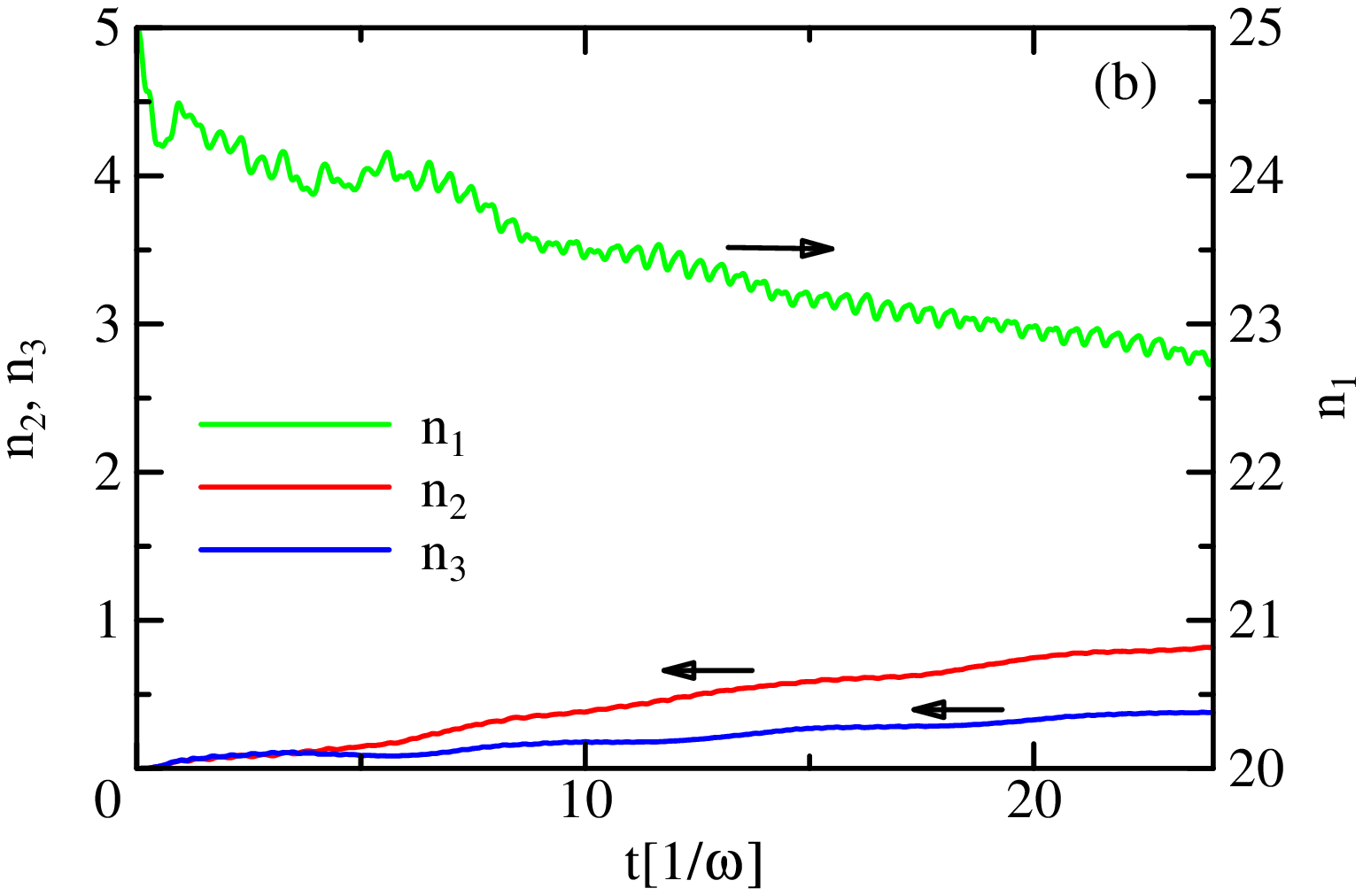}}
\caption{(a) The ground state population of electron $N(t)$ for $\lambda=1.5$ as functions of time. The solid red line shows the result for $n=3$ and $\alpha_2=\alpha_3=0$, and the blue line shows that for $n=1$. The dotted line is $N(t)$ by the semiclassical approximation for $n=3$. (b) photon number $n_i\ (i=1,2,3)$ for $n=3$ and $\alpha_2=\alpha_3=0$.}
\label{l15sas0}
\end{figure}
The solid red line in Fig.\ \ref{l15sas0}-(a) shows $N(t)$ for $n=3$ and $\alpha_2=\alpha_3=0$.
As the absorption of photon proceeds, $N(t)$ decreases and a rapid oscillatory behavior appears in $t \leq 2$.
This feature is a reminiscent of the Rabi oscillation though it diminishes with the lattice relaxation, i.e., emission of phonons.
When only a single-mode photon is taken into account, i.e., $n=1$, $N(t)$ behaves similarly to that for $n=3$, which shows that the Raman processes plays a minor role in the wavepacket dynamics.
This interpretation is also supported by the behavior of $n_2$ and $n_3$ shown in Fig.\ \ref{l15sas0}-(b).
This figure shows that the increase of the photon number for modes 2 and 3 is small even when the wavepacket motion proceeds and the energy difference between two corresponding adiabatic PESs is resonant to the Stokes mode.

\begin{figure}
\scalebox{0.5}{\includegraphics{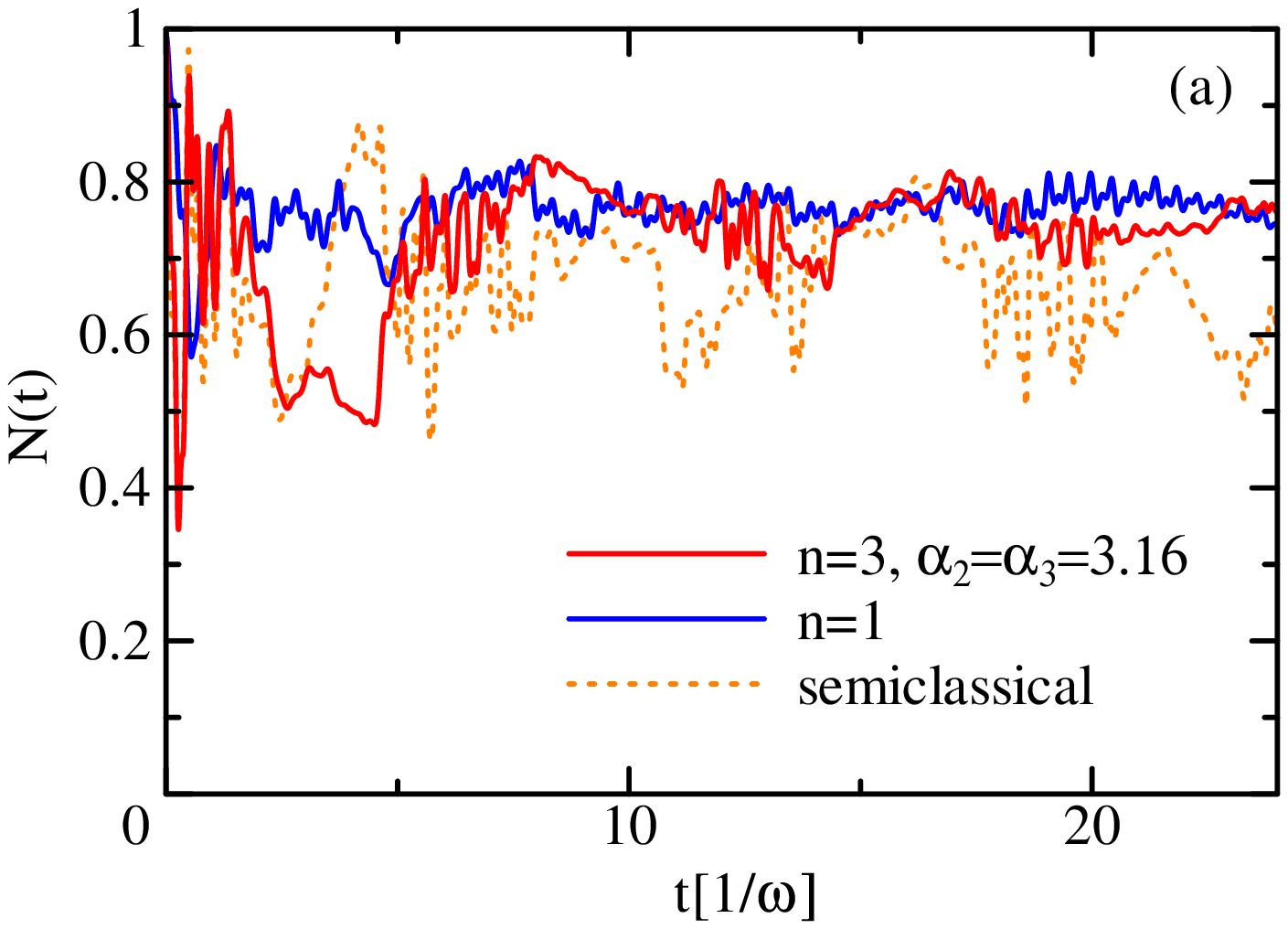}}
\scalebox{0.5}{\includegraphics{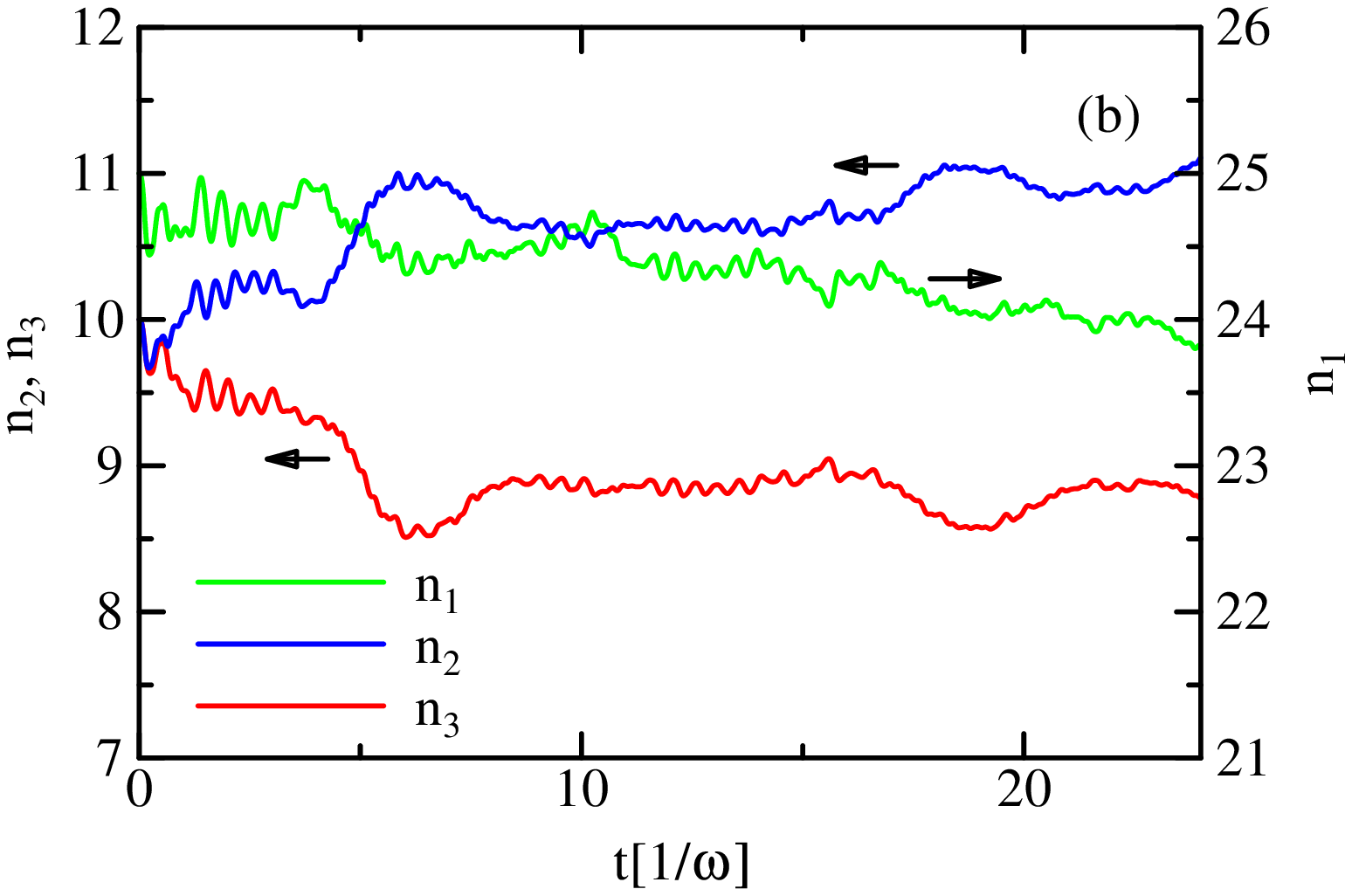}}
\caption{(a) The ground state population of electron $N(t)$ for $\lambda=1.5$ as functions of time. The solid line shows the quantum dynamics of $N(t)$ for $n=3$ and $\alpha_2=\alpha_3=3.16$, and the dotted line shows the result obtained by the semiclassical approximation for $n=3$. (b) photon number $n_i\ (i=1,2,3)$ for $n=3$ and $\alpha_2=\alpha_3=3.16$.}
\label{l15sas10}
\end{figure}
When both the Stokes and the anti-Stokes modes have finite intensity at $t=0$, time evolution of the system shows a different behavior.
The solid red line in Fig.\ \ref{l15sas10}-(a) shows $N(t)$ for $\alpha_2=\alpha_3=3.16$ and $\lambda=1.5$, and the dotted line is a corresponding property calculated by the semiclassical approximation.
Although both lines almost coincide with each other for $t < 2$, deviation between them increases rapidly thereafter.
Accordingly, as Fig.\ \ref{l15sas10}-(b) shows, photons in mode 2 and 3 increase and decrease, respectively, which shows that the Raman processes contribute to the dynamics of the whole system.
Precisely, $n_2$ and $n_3$ rapidly change their value at $t \sim 2\pi$, and the electronic transition is modulated by the Raman processes in this period of time.
In other words, the stimulated Raman process enhances the electronic transition for $\alpha_2 , \alpha_3 \neq 0$ and thus the wavepacket dynamics is significantly modulated.

Comparing the temporal behavior of $n_1$ in Figs.\ \ref{l15sas0}-(b) and \ref{l15sas10}-(b), we found that
the absorption of pump-mode photons is suppressed when both the Stokes and the anti-Stokes Raman processes take place.
Since these processes are relevant to both the absorption and the emission of the pump-mode photons, we consider that the interference between two Raman processes contributes to the dynamics of the whole system.
This effect does not disturb the the transition between $|g\rangle$ and $|e\rangle$ and thus $N(t)$ changes its value as Fig.\ \ref{l15sas0}-(a) shows.
Since such processes are not taken into account for $n=1$, the difference between the solid red line and the blue line in Fig.\ \ref{l15sas10}-(a) is much larger than that in Fig.\ \ref{l15sas0}-(a).

\begin{figure}
\scalebox{0.5}{\includegraphics{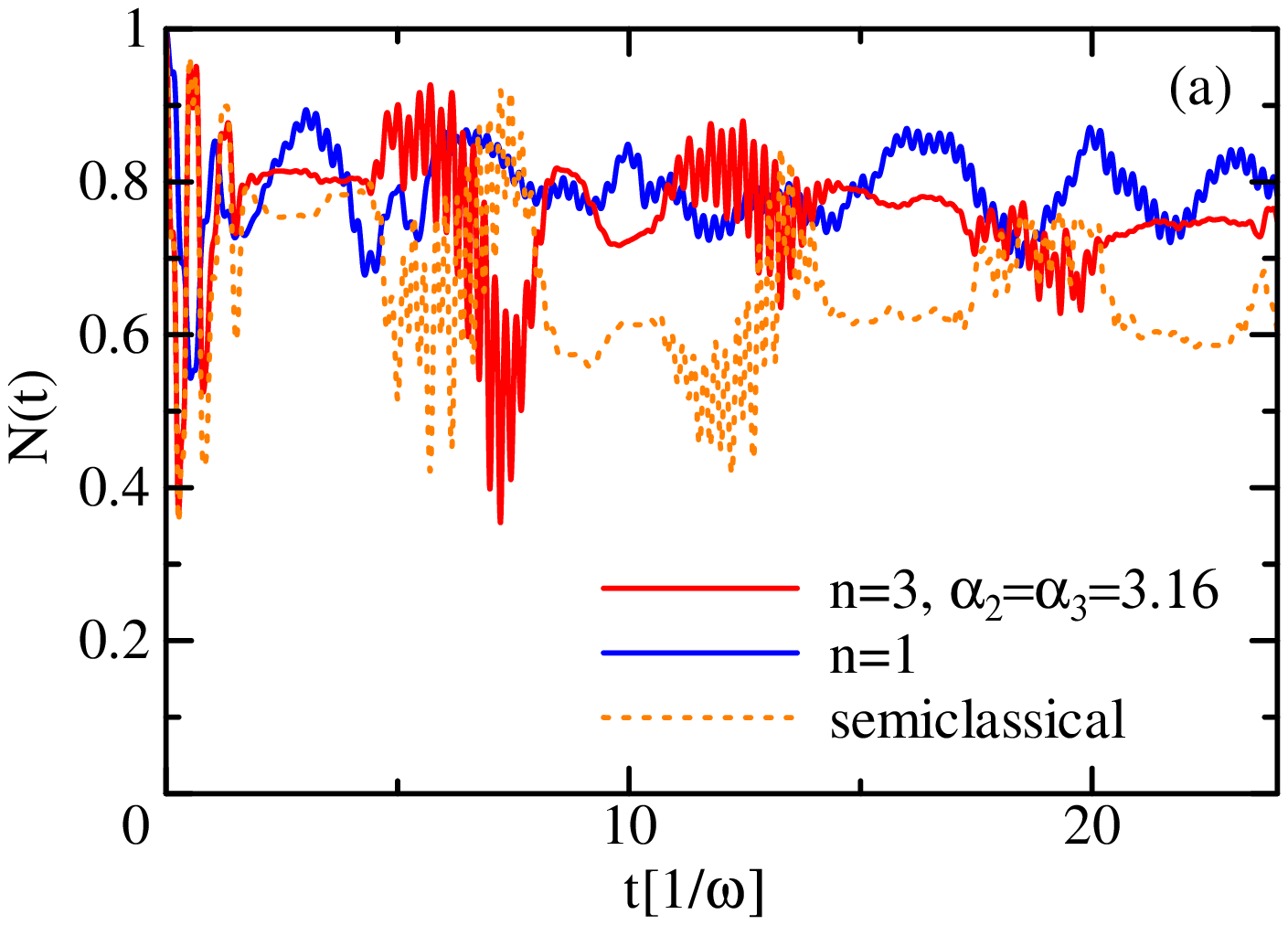}}
\scalebox{0.5}{\includegraphics{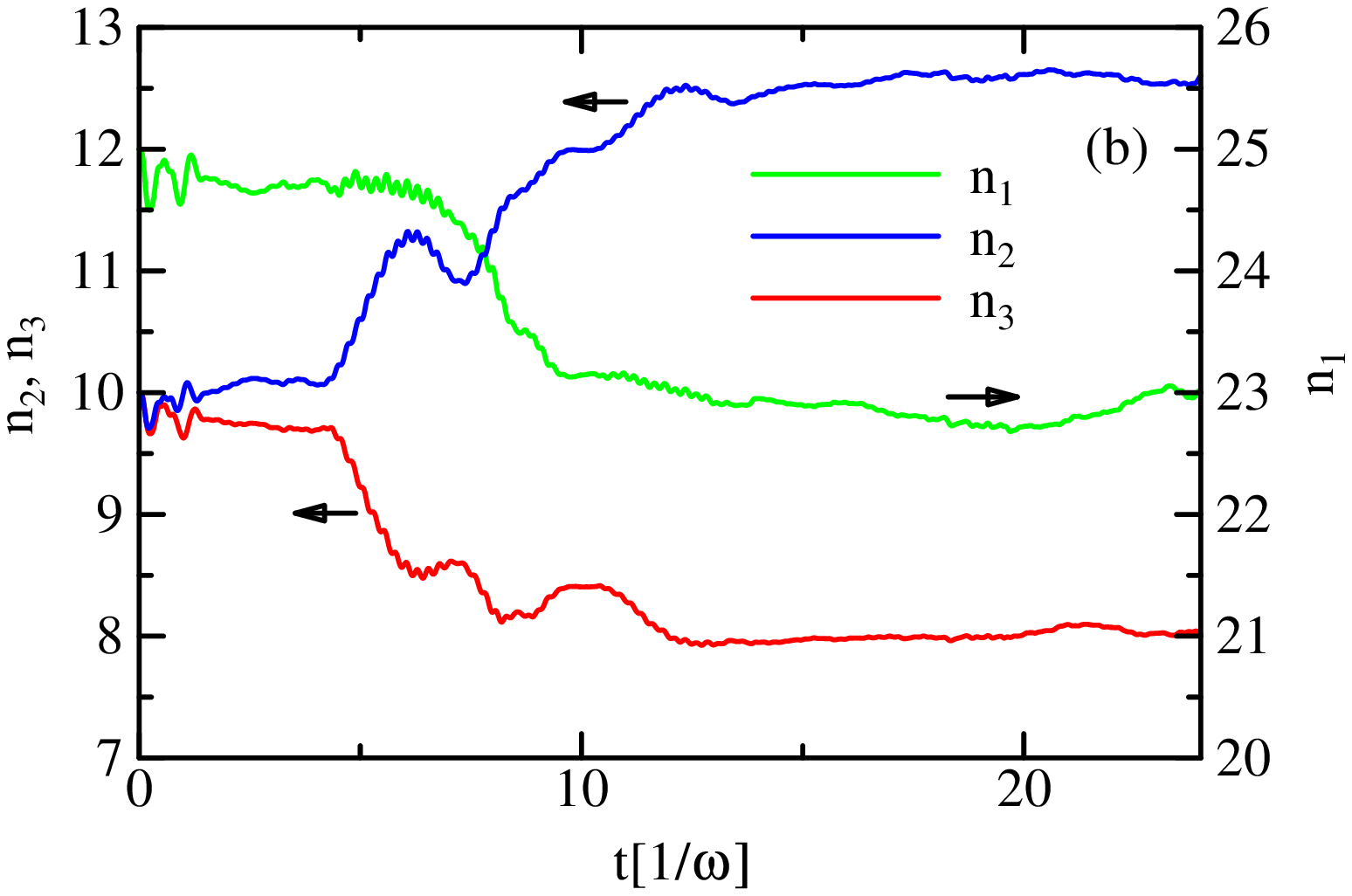}}
\caption{(a) The ground state population of electron $N(t)$ for $\lambda=0$ as functions of time. The solid line shows the quantum dynamics of $N(t)$ for $n=3$ and $\alpha_2=\alpha_3=3.16$, and the dotted line shows the result obtained by the semiclassical approximation for $n=3$. (b) photon number $n_i\ (i=1,2,3)$ for $n=3$ and $\alpha_2=\alpha_3=0$.}
\label{l0}
\end{figure}
These features are understood clearly by comparing the above results to those with the electron-phonon nonadiabatic coupling $\lambda$ turned off.
Figures \ref{l0}-(a) and (b) show $N(t)$ and $n_i(t)$ for $\alpha_2=\alpha_3=3.16$ and $\lambda=0$.
First, we point out the stepwise increase/decrease of $n_2$/$n_3$ around $t =2\pi$ and $4\pi$.
Furthermore, absorption of mode 2 and emission of mode 3 synchronously occur, while absorption of mode 1 takes place subsequently.
Since, however, no two-phonon/photon process is allowed to the lowest order, we point out that both Stokes and anti-Stokes Raman processes take place resonantly at this stage.
After the Raman processes become off-resonant, the interference between them is weakened and the absorption of mode 1 photons appears to be clearer.

Comparing Figs.\ \ref{l15sas10}-(a) with \ref{l0}-(a), we found that electronic transition is suppressed at $t \sim 2\pi$ and $4\pi$ also by the electron-phonon nonadiabaticity.
We note that the adiabatic PESs of the electron-phonon subsystem (PESSs) have an avoided crossing at $u = -\varepsilon/\nu'$ for $\lambda \neq 0$.
The photoexcited wavepackets bifurcate at the avoided-crossing and thus the transition between $|g\rangle$ and $|e\rangle$ is more complicated for finite $\lambda$.
In this case, the interference between those processes affects the electronic transition, and the resonance to the pump mode photons is blurred.
Hence, the temporal change of $N(t)$ and $n_i(t)$ for $t\sim 10$ is unclear as shown in Figs.\ \ref{l15sas10}-(a) and (b).

\begin{figure}
  \scalebox{0.5}{\includegraphics*{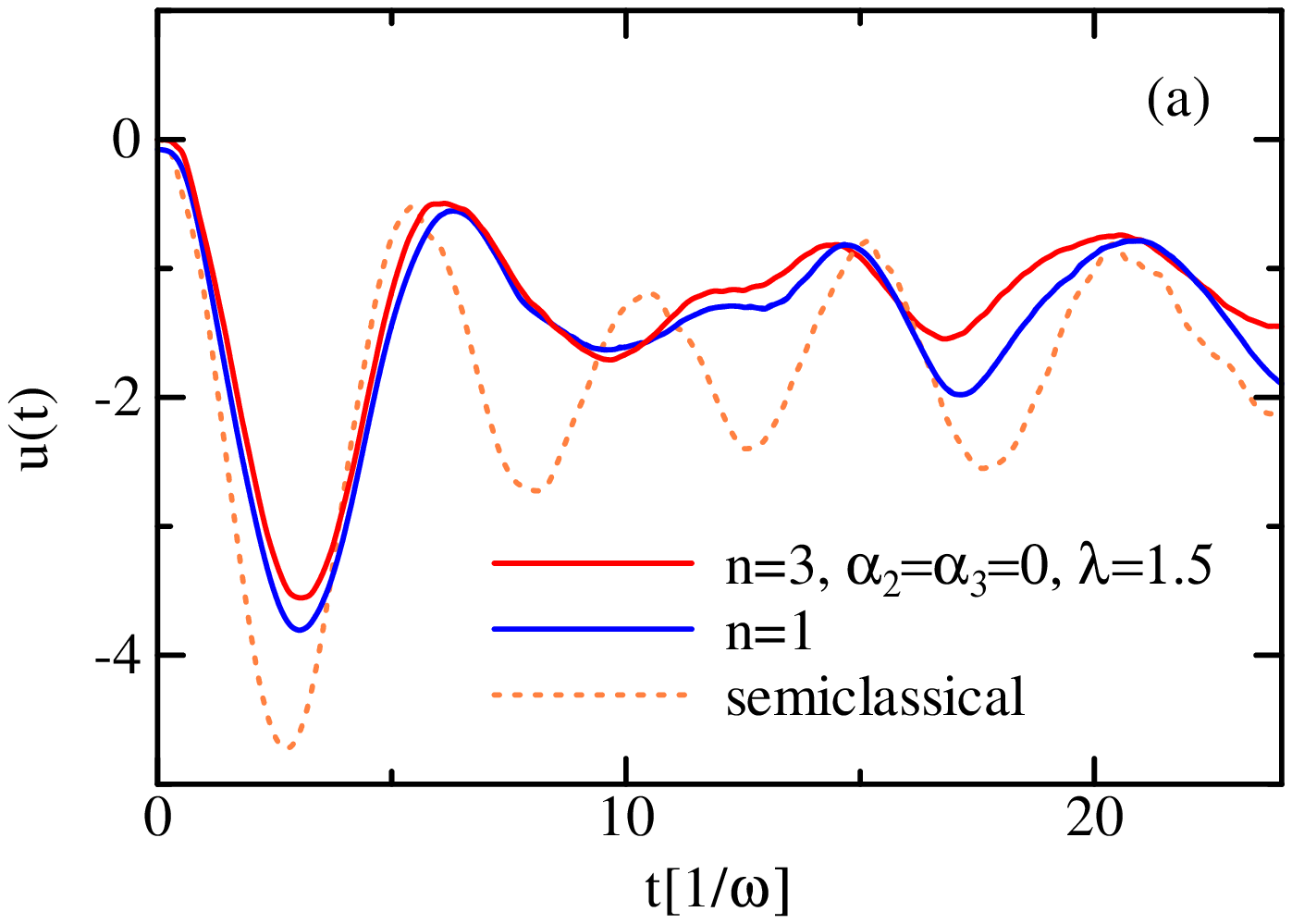}}
  \scalebox{0.5}{\includegraphics*{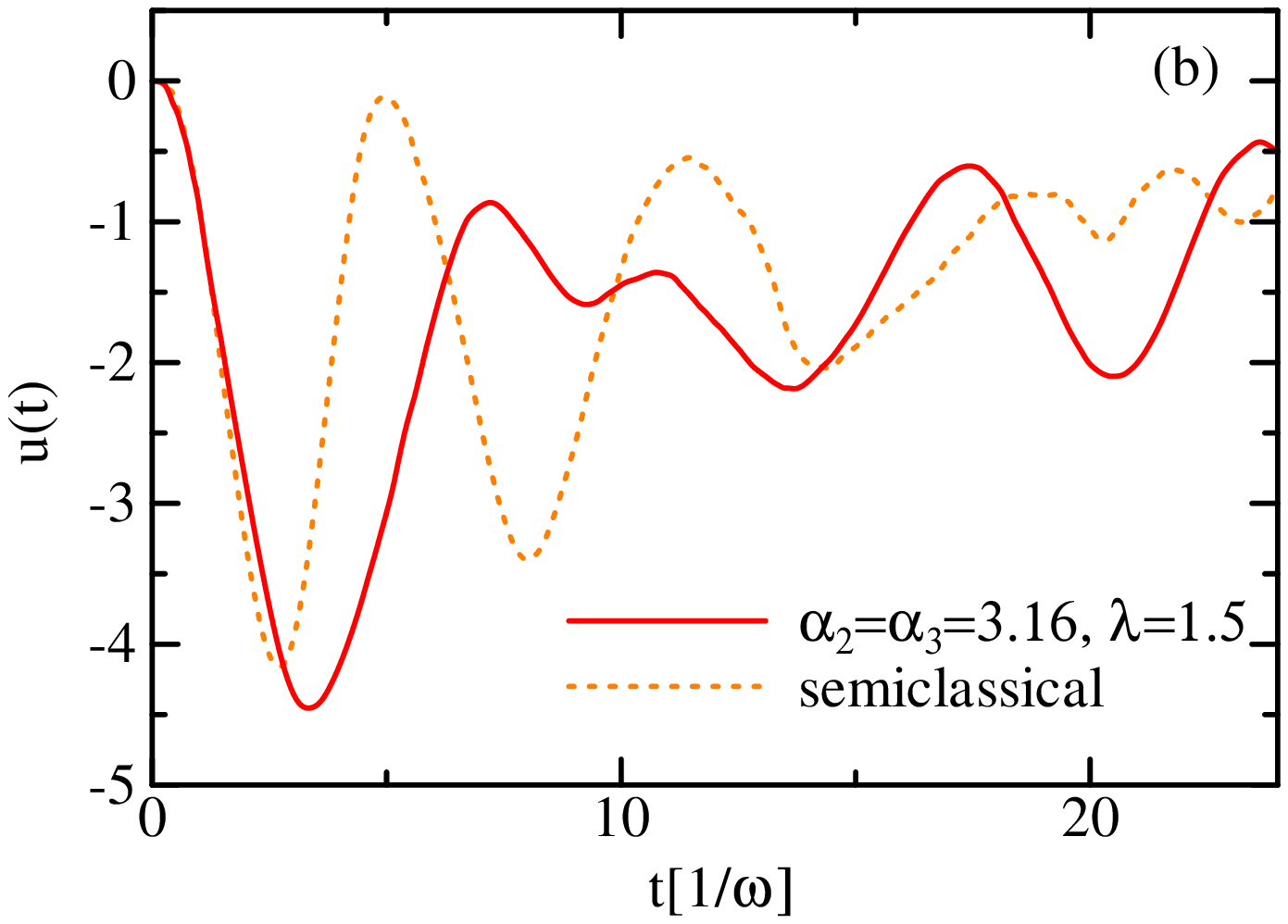}}
  \scalebox{0.5}{\includegraphics*{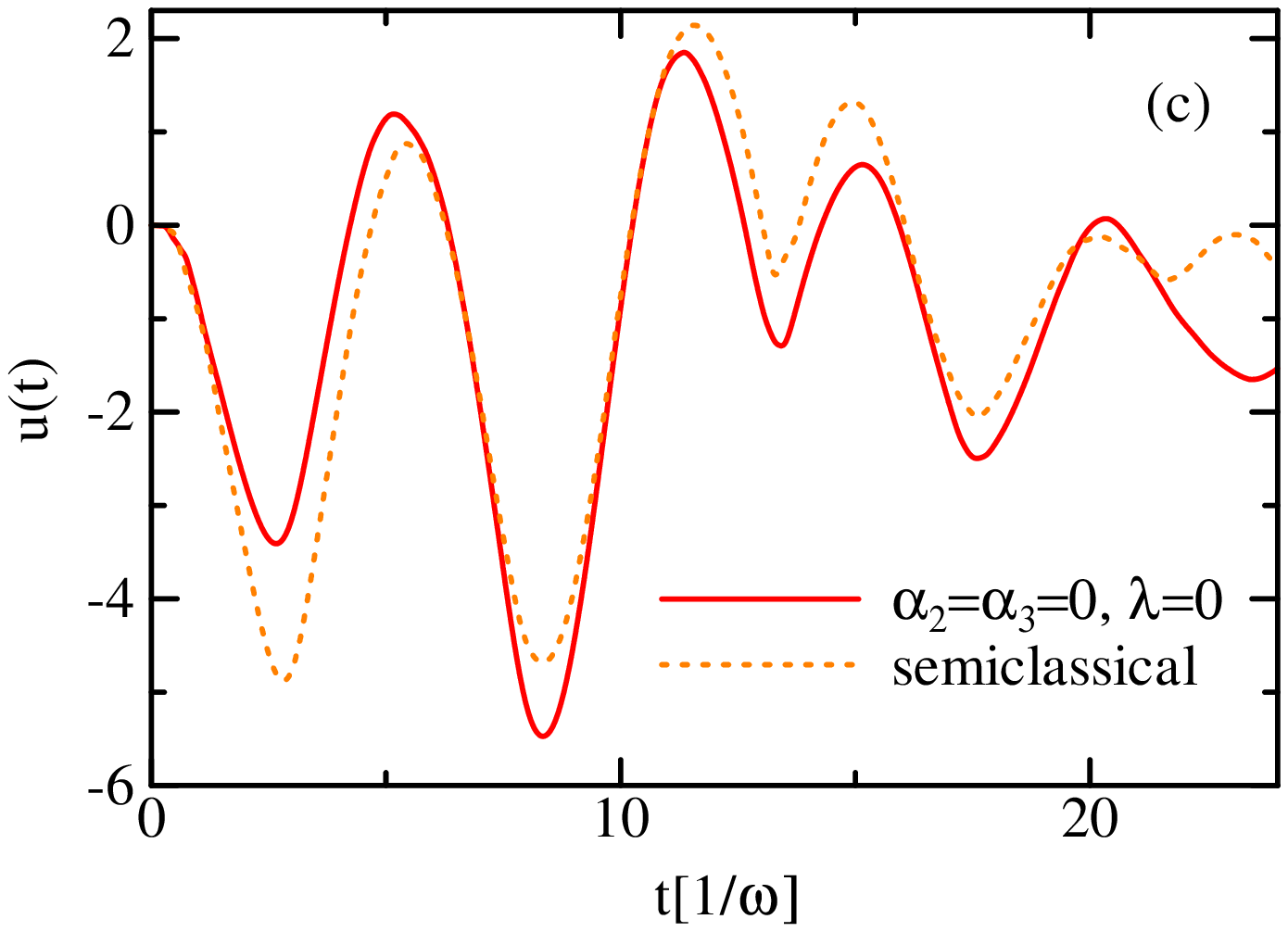}}
\caption{Lattice displacement $u(t)$ for (a) $\alpha_2=\alpha_3=0,\ \lambda=1.5$, (b) $\alpha_2=\alpha_3=3.16,\ \lambda=1.5$, and (a) $\alpha_2=\alpha_3=3.16,\ \lambda=0$. Calculated results with the semiclassical approximation are shown by the dotted line, and the blue line in (a) shows $u(t)$ for $n=1$.}
\label{disp}
\end{figure}
Since the wavepacket trajectory on the PESSs is an experimentally observable quantity\cite{ki2}, we calculated the lattice displacement $u(t)=\langle \Phi(t) |\hat{u}| \Phi(t) \rangle$ and focus on the dynamics of the electron and phonons.
Figures \ref{disp}-(a)-(c) show $u(t)$ for the three cases corresponding to Figs.\ \ref{l15sas0}-\ref{l0}.
As discussed previously,  Fig.\ \ref{disp}-(a) and (b) show that the semiclassical approximation is not valid for $t > 2\pi$ for $\lambda \neq 0$.
On the contrary, the solid line and the dotted line in Fig.\ \ref{disp}-(c) are similar to each other, which shows that the nonadiabaticity of the electron-phonon-photon dynamics is relevant to the validity of the semiclassical approximation.
As mentioned above, the (one-dimensional) PESSs have an avoided-crossing at $u = -\varepsilon/\nu'$, while the PESs for the whole system has a CI.
Since the semiclassical approximation takes into account only the avoided-crossing, the wavepacket motion bifurcate in a different manner between the quantum-mechanical calculation and the semiclassical calculation particularly in the vicinity of the avoided-crossing or the CI, which results in the different dynamics or trajectory shown in the figures.
We stress that the quantum-mechanical nature of the incident light plays an important role on the wavepacket motion, and that detailed discussion will be possible by revealing the transient dynamics of coherent phonons by ultrafast optical spectroscopy.
To be more precise, the role of the CI on the electronic transition should be revealed in order to determine the wavepackets created by photons.
In particular, as the irradiation of photons proceeds, deviation from single-mode model becomes larger, the role of CI becomes more important.

\begin{figure}
  \scalebox{0.5}{\includegraphics*{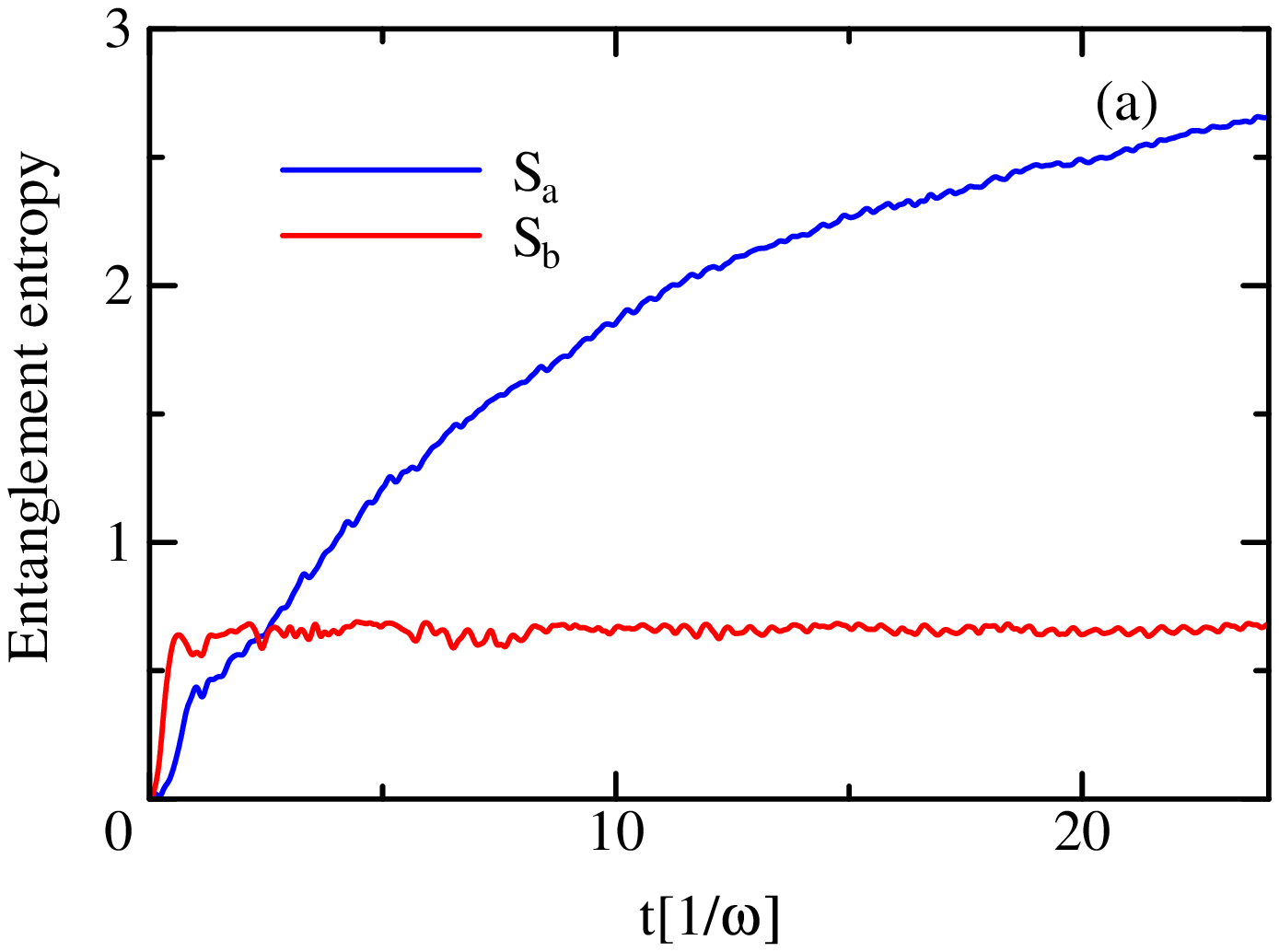}}
  \scalebox{0.5}{\includegraphics*{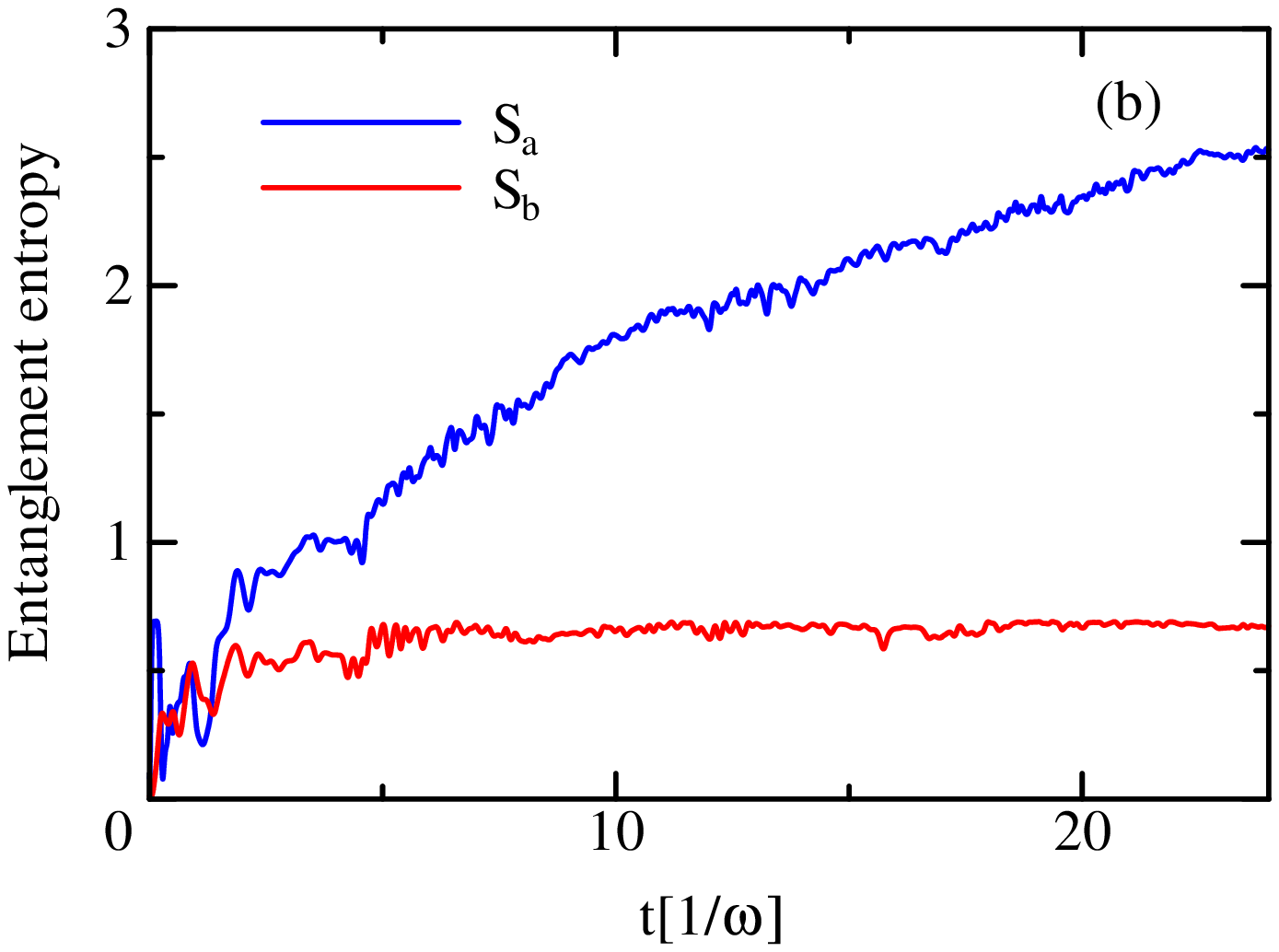}}
  \scalebox{0.5}{\includegraphics*{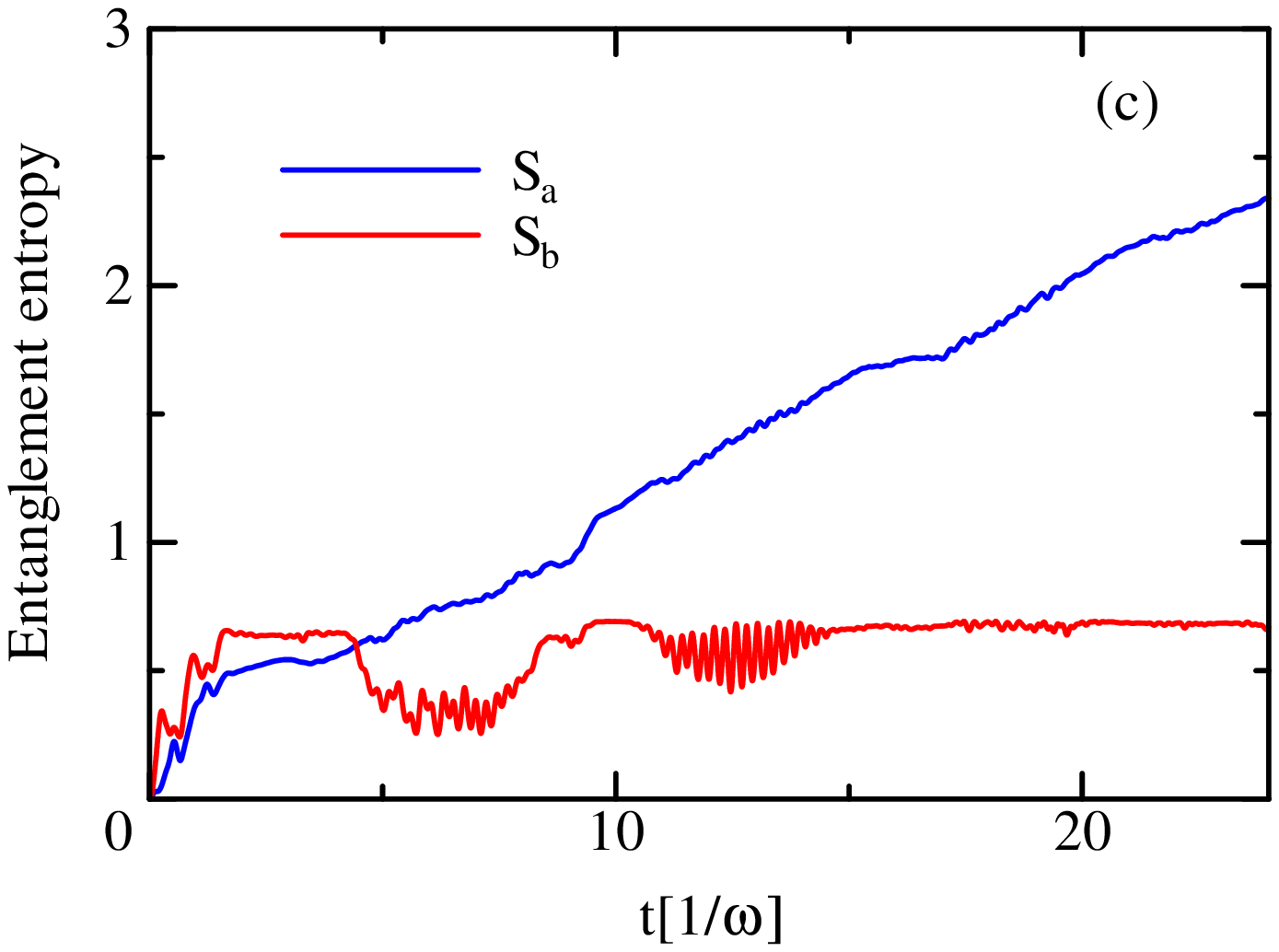}}
\caption{Entanglement entropy $S_a(t)$ and $S_b(t)$ for (a) $\alpha_2=\alpha_3=0,\ \lambda=1.5$, (b) $\alpha_2=\alpha_3=3.16,\ \lambda=1.5$, and (c) $\alpha_2=\alpha_3=3.16,\ \lambda=0$.}
\label{entropy}
\end{figure}
The quantum-mechanical nature of the electromagnetic field is reflected on the entanglement between subsystems, i.e., photons, phonons, and electrons.
In this paper we calculated the bipartite entanglement entropy in which the whole system is divided into (i) the electronic system and the phonon-photon system, and (ii) the electron-phonon subsystem and the photons.
We calculated the entropy for these cases defined by
\begin{eqnarray}
  S_a(t) & = & {\rm Tr} \rho_a(t) \log \rho_a(t),\\
  S_b(t) & = & {\rm Tr} \rho_b(t) \log \rho_b(t),
\end{eqnarray}
where
\begin{eqnarray}
  \rho_a(t) & = & {\rm Tr}_{pn,pt} |\Phi(t)\rangle \langle\Phi(t)|,\\
  \rho_b(t) & = & {\rm Tr}_{pt} |\Phi(t)\rangle \langle\Phi(t)|.
\end{eqnarray}
Tr$_{pn,pt}$ and Tr$_{pt}$ denote the partial trace of the density matrix regarding the phonon and photon degrees of freedom, and the photon degrees of freedom, respectively.
We point out that $S_a$ is the entropy of a two-level system and that its value lies between 0 and $\log 2 \sim 0.693$.

Figures \ref{entropy}-(a)-(c) show that entanglement represented by (i) and (ii) grows immediately after the simulation starts.
Figures \ref{entropy}-(b) and (c) also show that $S_a$ and $S_b$ have a fine structure corresponding to the increase/decrease of $n_2$/$n_3$, i.e., the Raman processes enhance the rate of entropy production.
As for the effect of the electron-phonon nonadiabaticity, we point out that Fig.\ \ref{entropy}-(c) shows that the photoabsorption at $t \sim 2\pi$ and $4\pi$ decreases $S_b$, which shows that the coherence between the electronic states recovers by the external field, i.e., photons.
As shown in the other properties, the interference mechanism between electronic states becomes different in the presence of the electron-phonon nonadiabaticity and the decrease of $S_b$ is not observed in Figs.\ \ref{entropy}-(a) or (b).

These figures also show that the quantum information carried in the electron-phonon system is able to be transferred to another material system by irradiated photons.
However, entanglement between photons and phonons is also disturbed by their interaction with electrons through $\mu_i$ and $\nu$, and further discussion on the entanglement entropy in multipartite systems\cite{multi1,multi2,multi3} will be helpful to clarify the entanglement properties of the system.

\section{Conclusions}
\label{concl}
We calculated the wavepacket dynamics of electron-phonon-photon systems by numerical calculations focusing on the interplay of the electron-phonon nonadiabaticity and the Raman processes.
While the effect of the Raman processes is small on the wavepacket motion starting with empty Stokes/anti-Stokes modes, the stimulated Raman processes enhance the electronic transition, which shows that we should take into account them to obtain accurate trajectories of wavepackets.
In particular, the interference between Stokes and anti-Stokes Raman processes suppresses the absorption of pump-mode photons, although the electronic excitation is induced and the excited-state population $N(t)$ takes a large value.
Furthermore, when both $\lambda$ and $\nu_i$ are present, large deviation from semiclassical calculations is obtained, which shows that the bifurcation of wavepackets should be carefully treated in the presence of both the electron-phonon adiabaticity and the Raman processes.

Entanglement between the electron-phonon system and the photons shows that this system is applicable to the quantum information technology, though the detailed discussion on the multipartite entanglement entropy is necessary.

We have not shown a direct evidence on the role of the CI in the present study.
The CI acts as a ``magnetic flux'' in the Aharanov-Bohm effect, and thus the geometric properties of the wavefunction on multidimensional PESs has been studied\cite{ci2,citheory,xie}.
We consider it helpful to reveal the detail of quantum nature of the photoexcited state as well as the wavepacket dynamics of coupled electron-phonon-photon systems by a similar method.

The author is grateful to K. G. Nakamura for fruitful discussion.
Numerical calculations were done using the facilities of the Supercomputer Center, the Institute for Solid State Physics, the University of Tokyo, Japan.

\end{document}